\documentstyle[12pt]{article}

\begin{document}


\def\a{\alpha}
\def\b{\beta}
\def\c{\varepsilon}
\def\d{\delta}
\def\e{\epsilon}
\def\f{\phi}
\def\g{\gamma}
\def\h{\theta}
\def\k{\kappa}
\def\l{\lambda}
\def\m{\mu}
\def\n{\nu}
\def\p{\psi}
\def\q{\partial}
\def\r{\rho}
\def\s{\sigma}
\def\t{\tau}
\def\u{\upsilon}
\def\v{\varphi}
\def\w{\omega}
\def\x{\xi}
\def\y{\eta}
\def\z{\zeta}
\def\D{\Delta}
\def\G{\Gamma}
\def\L{\Lambda}
\def\F{\Phi}
\def\P{\Psi}
\def\S{\Sigma}

\def\o{\over}
\def\beq{\begin{eqnarray}}
\def\eeq{\end{eqnarray}}
\newcommand{\gsim}{ \mathop{}_{\textstyle \sim}^{\textstyle >} }
\newcommand{\lsim}{ \mathop{}_{\textstyle \sim}^{\textstyle <} }

\def\IJMP{Int.~J.~Mod.~Phys. }
\def\MPL{Mod.~Phys.~Lett. }
\def\NP{Nucl.~Phys. }
\def\PL{Phys.~Lett. }
\def\PR{Phys.~Rev. }
\def\PRL{Phys.~Rev.~Lett. }
\def\PTP{Prog.~Theor.~Phys. }
\def\ZP{Z.~Phys. }


\baselineskip 0.7cm

\begin{titlepage}
\begin{flushright}
UT-843
\\
April, 1999
\end{flushright}

\vskip 1.35cm
\begin{center}
{\large \bf 
Dynamical Inflation and Vacuum Selection
}
\vskip 1.2cm
Izawa K.-I. and T.~Yanagida
\vskip 0.4cm

{\it Department of Physics, University of Tokyo,\\
     Tokyo 113-0033, Japan}

{\it Research Center for the Early Universe, University of Tokyo,\\
     Tokyo 113-0033, Japan}

\vskip 1.5cm

\abstract{
We consider an inflationary scenario
where the energy scale of
inflation stems from gauge theory dynamics.
We point out its generic implications on vacuum selection of our universe,
in particular, on determination of spacetime symmetries
including its dimensions.
}
\end{center}
\end{titlepage}

\setcounter{page}{2}


The energy scale of primordial inflation is supposed to be
hierarchically below the gravitational scale in order to produce
tiny fluctuations of the cosmic microwave background radiation
\cite{Lyt}.
The origin of the hierarchical scale of inflation
may be dimensional transmutation induced by gauge theory dynamics.%
\footnote{This is analogous to dynamical supersymmetry breaking,
which may provide the origin of the hierarchically small scale of
electroweak symmetry breaking.}   
The gauge theory dynamics for inflation is expected to cause profound
effects on the evolution of the universe, since
the inflationary process may determine the vacuum of our universe
through dominant expansion of the spatial extension of the corresponding
inflationary vacuum.

In this paper,
we consider implications of this dynamical inflation
on vacuum selection of our universe,
in particular, on determination of spacetime symmetries
including its dimensions.

Now we expose a mixed model of
dilaton fixing
\cite{Iza}
and topological inflation
\cite{Kaw}%
\footnote{They are both $R$-invariant and so is the mixed model.}
as a simple example of dynamical inflation
which incorporates an inflationary selection of the vacuum
of our universe.

Let us consider a four-dimensional supersymmetric field theory
with a dilaton $\F$ and an inflaton $\f$ supermultiplets
whose superpotential takes a form
\beq
 W = Xf(\F) + Zh(\F)(1 - \l \f^2),
 \label{SP}
\eeq
where $X$ and $Z$ denote chiral superfields
and $f(\F)$ and $h(\F)$ are functions of
the dilaton superfield $\F$:
\beq
 f(\F) = f_1 e^{-a_1 \F} - f_2 e^{-a_2 \F}, \quad
 h(\F) = h e^{-a \F}.
\eeq
Here we set the gravitational scale equal to unity and regard it
as a universal cutoff in the theory.
We assume the couplings $f_1$, $f_2$, $h$ and $\l$ of order one and
 $0 < a_1 \sim a_2 \ll a$.
This superpotential can be generated by dynamics of hypercolor
gauge interactions where the dilaton-dependent scales $e^{-a_i \F}$
and $e^{-a \F}$ with $a$'s related to $\b$ functions of the gauge
interactions
arise from hyperquark condensations with the aid of a superpotential
considered in Ref.\cite{Iza}.

The potential in supergravity is given by
\begin{equation}
 V = e^{K} (K_{AB} F^A F^{B*} - 3|W|^2),
 \label{POT}
\end{equation}
where $K$ is a K{\" a}hler potential,
$K_{AB}$ denotes the inverse of the matrix
\beq
 {\q^2 K \o \q \f_A \q \f_B^*},
\eeq
with $\f_A = X, Z, \F, \f$,
and $F^A$ is given by
\beq
 F^A = {\q W \o \q \f_A} + {\q K \o \q \f_A}W.
\eeq

For a generic K{\" a}hler potential, we have vacua of the model
given by
\beq
 F^A = W = 0,
 \label{COND}
\eeq
which realizes the unbroken supersymmetry and $R$ invariance.
Hence the vacuum expectation values of the dilaton are determined by
\beq
 f(\F) = 0, 
\eeq
which has runaway and fixed solutions,
\beq
 \F \rightarrow \infty
\eeq
and
\beq
 \langle \F \rangle = {1 \o a_1-a_2}\ln {f_1 \o f_2},
\eeq
respectively.%
\footnote{The gauge coupling constant $g$ is given by
$g^2 = 1/{\rm Re}\langle \F \rangle$.}

For the fixed dilaton, from Eq.(\ref{COND}), we find
$\langle X \rangle = \langle Z \rangle =0$
and
\beq
 \langle \f \rangle = \pm \l^{-{1 \o 2}}.
 \label{VAC}
\eeq
Since $|e^{-a_i \langle \F \rangle}| \gg |e^{-a \langle \F \rangle}|$,
we may integrate out the superfields $X$ and $\F$ to obtain
an effective superpotential
\beq
 W_{eff} = Zh(\langle \F \rangle)(1 - \l \f^2),
\eeq
which results in topological inflation between the two vacua
Eq.(\ref{VAC}) for appropriate values of the couplings
\cite{Kaw}.
Namely, once the dilaton is fixed, the universe may evolve through
an inflationary stage. 

On the other hand, the runaway vacuum
$\F \rightarrow \infty$
yields a free theory, which induces no inflation.

Under the chaotic initial condition
\cite{Lin}
of the dilaton field, the spatial extension of the vacuum
corresponding to the fixed dilaton dominates through the inflationary
process over that of the runaway vacuum
\cite{Iza}.

Although we have provided a specific model, for definiteness,
to demonstrate our point, the implications we consider
in the dynamical inflation scenario seem rather generic. 
In the above model, the dilaton serves as an example of moduli%
\footnote{For an investigation of stringy modular cosmology, see
Ref.\cite{Ban}.}
which determine the form of low-energy field theory
describing our universe.
Moduli are not necessarily usual moduli fields but
may be any variables which parametrize the moduli space of
underlying high-energy theory.
The moduli space may be even disconnected, since the chaotic
initial condition of moduli allows highly excited states
which interpolate disconnected pieces of the moduli space. 

For example, the spacetime dimension of the low-energy field theory
may be determined by moduli which describe the size of the internal
space of high-energy theory. Then the chaotic initial condition
of those moduli implies that the spacetime dimension is determined
through the inflationary process. If the corresponding inflation
is dynamical, the spacetime dimension is expected to be four or less,
since the dynamical scale due to gauge theory dynamics
may not be generated in five or more dimensions. 

In four dimensions,
the presence of ${\cal N}=1$ supersymmetry (and not ${\cal N} > 1$)
may also originate from
inflationary selection of this type, since it seems suitable for
realizing scalar potentials which satisfy the slow-roll condition
for inflation
\cite{Lyt}.


\newpage

\begin{thebibliography}{99}

\bibitem{Lyt}
  For a review, D.H.~Lyth and A.~Riotto, hep-ph/9807278.

\bibitem{Iza}
  Izawa~K.-I. and T.~Yanagida, \PTP {\bf 101} (1999) 171,
  hep-ph/9809366.

\bibitem{Kaw}
  Izawa~K.-I., M.~Kawasaki and T.~Yanagida, hep-ph/9810537.

\bibitem{Lin}
  A.D.~Linde, {\sl Particle Physics and Inflationary Cosmology},
  (Harwood, 1990).

\bibitem{Ban}
  T.~Banks, M.~Berkooz, S.H.~Shenker, G.~Moore and P.J.~Steinhardt,
  \PR {\bf D52} (1995) 3548, hep-th/9503114; \\
  T.~Banks, hep-th/9601151.

\end{thebibliography}
\end{document}